\documentstyle[11pt]{article}
\input{psfig}
\begin{document}
\begin{flushright}
Liverpool Preprint: LTH 321\\
hep-lat/9310026\\
October 19, 1993
\end{flushright}
\vspace{5mm}
\begin{center}
{\LARGE\bf
Fitting Correlated Data
}\\[1cm]

{\bf C. Michael}\\

\it{DAMTP, University of Liverpool, Liverpool, L69 3BX, U.K.}

\end{center}

\begin{abstract}
We discuss fitting correlated data - with the example of
hadron mass spectroscopy in mind. The main conclusion is that
the method of minimising correlated $\chi^2$ is unreliable
if the data sample is too small.
\end{abstract}

\section{Specification of the problem}

We have $N$ samples of unbiased estimators of
quantities $x_i$ with $i=1 \dots D$. Thus the data set is
$x_i(n)$ where $n=1 \dots N$. We assume that the samples
$x_i(n)$ are statistically independent versus $n$ for
fixed $i$ but may be correlated in $i$.
Such a situation arises in lattice gauge theory calculations
where there are $N$ independent configurations and $D$
Green functions (linear combinations of Wilson loops or
propagators) are measured versus time separation $i$. An
introduction to this topic in the context of
lattice gauge theory is provided by Toussaint~\cite{T}.

The aim is to fit a given function $F_i$ which depends on $P$
parameters $a_p$. This function is to be fitted to the
data samples $x_i$.   Thus we require to find the following
\begin{itemize}
\item the best values of the parameters $a_p$.
\item the errors associated with these best fit parameters.
\item the probability that the fit represents the data sample.
\end{itemize}

\section{Representation of the probability distribution of the sample}

The data sample themselves give a probability distribution
$$
S(x)={1 \over N} \sum_{n=1}^{N} \delta^D (x-x(n))
$$
We shall be interested in estimates of the probability distribution
of the averages $X_i$ of the data $x_i$.
 The most general way to achieve this
is to fold the above distribution $N$ times
$$
Q_S(X)= \int \delta^D (X- {1 \over N} \sum_{r=1}^N x^{(r)})
 \prod_{s=1}^N    d^Dx^{(s)} S(x^{(s)})
   $$
 Evaluating this
distribution by simulation corresponds to the  bootstrap method:  many
samples of N data are obtained by choosing  from the original N
possibilities  randomly (with repetition  allowed). In the limit of many
such samples this corresponds  to the above distribution $Q_S$.  Such a
procedure is in general  inadequate for determining best fit parameters
since a smooth representation  of $Q_S(X)$ is needed.
An even  more difficult task is usually to estimate the acceptability
of such a  fit. Thus the best fit parameters will yield
$X(f)$ and one must estimate the probability of such a value
arising stochastically. This also needs a smooth model of
the distribution $Q_S$.

The natural interpolation  for $Q_S$ is suggested by the central limit
theorem. Provided the underlying distributions of $x_i$ are
sufficiently localised, then for large $N$, $X_i$ will be
gaussianly distributed.  We are specifically interested in the
case where the different components $x_i$ are statistically
correlated. Thus a general gaussian surface will be needed.
$$
Q_G(X)= H \exp( -{1 \over 2}
(X_i - \overline{X}_i) M_{ij} (X_j - \overline{X}_j) )
$$
Equating the first and second moments of these two expressions
leads to the well known identifications
$$
\overline{X}_i={1 \over N} \sum_{n=1}^N x_i
$$
$$
M_{ij}=N C^{-1}_{ij} \ \hbox{ ,where}
$$
$$
C_{ij}=    {1 \over N-1} \sum_{n=1}^N
(x_i-\overline{X}_i)(x_j-\overline{X}_j)
$$
To find the best fit parameters then corresponds to maximising
$$
\exp(-\chi^2 /2) \ \hbox {where } \
 \chi^2= (F_i(a)-\overline{X}_i) M_{ij} (F_j(a)-\overline{X}_j)
$$
with respect to $a_p$ for $p=1 \dots P$. This is the usual
correlated $\chi^2$ method.

Diagonalising the real symmetric positive-definite matrix $C$,
then allows us to write
$$
Q_G = H \exp (- \chi^2/2) =H \exp (-\sum_k^D \chi^2_k/2)
    = H \exp ( - \sum_k^D (Y_k-\overline{Y}_k)^2 /2 )
$$
$$
\hbox{where } \  C_{ij}=R_{ki} \lambda_k R_{kj},
\ \overline{Y}_j=R_{ji} \overline{X}_i N^{1 \over 2}/\lambda_i^{1 \over 2} \
 \hbox{ and }  \ Y_j=R_{ji}X_i N^{1 \over 2} / \lambda_i^{1 \over 2}.
$$
Consider, for example, a zero-parameter fit to $X_i=F_i$
where $F_i$ is the true
value (the average over many samples). Then for one sample,
the maximum is at $X_i=\overline{X}_i \ne F_i$
 and one can estimate the expected value of $\chi^2$.
Now letting ${\cal F}_i = R_{ij} F_j N^{1 \over 2} / \lambda_j^{1 \over 2}$,
 the probability distribution
$Q_G$ is a sum of $D$ independent variables $(Y_k-{\cal F}_k)$ with
unit variance.  Thus $<\chi^2>=D$ is expected.
As we shall see later,
there may be quite large corrections to this estimate in practice.

For a general fit $X_i=F_i(a)$ with $P$ parameters $a$, it is useful
to regard this as a constraint on the uncorrelated variables
$Y_k={\cal F}_k(a)$. The set of ${\cal F}_k$
 values as the parameters $a$ vary will
be a $P$-dimensional surface and the fit minimises $\chi^2$ within this
surface. So effectively $\chi_k^2$ is zero for $P$ components and
thus the expected value of $\chi^2$ only comes from
the remaining $D-P$ dimensions or \lq degrees of freedom\rq{}.
Thus in the simplest case one expects $\chi^2$ to be $D-P$.

What has become well known in the Lattice Gauge Theory community is
that the correlated $\chi^2$ approach
 is not a very stable method~\cite{P}.  Examples can also  be
constructed which give counter intuitive results~\cite{S}.
 I shall illustrate this with a simple simulation.

\section{A spherical gaussian distribution}

For illustration, consider a true probability distribution of data points
$$
S_T(x)= J \exp(-x_i x_i /2)
$$
where $i=1 \dots D$. This is actually quite general since after
a change of variables (translation, rotation and scaling),
 a general gaussian surface can be brought to this form.
 Then we take $N$ samples from this distribution
giving data $x_i(n)$ for $n=1 \dots N$. We obtain the distribution
$Q_G$ for this sample. We can easily maximise this  distribution
and clearly the sample maxima
will be at $X_i = \overline{X}_i$.  However the true maxima is
at $X_i=0$. Thus we can estimate the probability of a particular
sample having a maximum value of $X_i$. As well as using the
above \lq correlated $\chi^2$\rq{} approach, we also use a
diagonal approximation for comparison. Thus we can calibrate
standard statistical tests (such as $\chi^2$ per degree of
freedom) with this simulation in which the exact result is known.

Consider first the diagonal $\chi^2$ for orientation. Then we  evaluate
by simulation the average value of $\chi^2$ for the  fit to $X_i=F_i=0$.
This corresponds to a zero parameter fit and so it is only  the goodness
of fit that can be addressed. This is usually  quantified in terms of
the  value  of the exponent $\chi^2$. The model and analysis then treat
each  of the $D$ directions independently, so we obtain a value  of
$\chi^2 /D$ independent of $D$.
 The value of  $\chi^2=N \overline{X}_i^2/C_{ii}$ for each dimension. Now
the averages of $\overline{X}_i^2$ and $C_{ii}$ are $1/N$ and 1 respectively,
with both these estimates being unbiased. Thus if fluctuations
among samples for both these quantities are neglected,  we
obtain an  average of $\chi^2/D$ of 1 as expected.
However, their ratio
will be a biased estimator in general.
For small $N$ values, because the
 sample values fluctuate  quite widely (the variance of
$C_{ii}$ is $2/(N-1)$), the value of  $\chi^2= \overline{X}_i^2/C_{ii}$
for each dimension is enhanced on average by the contribution from
low values of
$C_{ii}$. Indeed for large $N$, analytic evaluation gives $\chi^2/D=1+2/N$.
 Our simulation values are shown in Table~\ref{tabletwo}
as the $D=1$ column and are consistent
with this analytic estimate.  Thus there can be a
small bias in the expected value of $\chi^2/D$ even for
uncorrelated evaluations.
As we shall see such biases become much stronger for
correlated $\chi^2$.

The way to illustrate the problem is to look at the shape
of the gaussian surface given from the sample data.
One way to visualise the shape of the sample distribution
$Q_G$ is to obtain the eigenvalues $\lambda$ and
eigenvectors of $C$. Small eigenvalues correspond to a
narrow distribution in the direction of the appropriate
eigenvector. The width in any direction of the
true distribution is 1. What we find by simulation is that
the smallest eigenvalue of $C$ from a sample of $N$ data points can
be very much smaller than 1 for small $N$. Of course,
for $N=D$, the points will  be linearly degenerate and
thus $\lambda_{min}=0$.  The results are shown in fig.~1 and
Table~\ref{tableone}.
One sees that for $N=55$ (a typical number of configurations
in a hadron spectrum calculation) and for $D=7$ the
smallest eigenvalue is around 0.5 on average. Thus the
distribution is 70\% narrower in the direction
corresponding to that eigenvalue.

The consequences of such a deformation (ie that the underlying
spherical distribution appears squashed in the sample)
are considerable in attempting to fit to the sample
distribution. For example, let us consider again the fit
of the true value $X_i=0$ to the sample distribution $Q_G$.
This corresponds to a zero parameter fit and so it is only
the goodness of fit that can be addressed:
 $\chi^2$ is expected to have an
average value of $D$. Our results are shown in Table~\ref{tabletwo}.
For example, we find,
for the correlated fit,  $\chi^2/D$
 is 3.03 with $D=5$ and $N=10$. Using the usual criterion of
$\chi^2$ per degree of freedom would then imply that there
was a very small probability (about 1\%) that $X_i=0$ was the maximum
of the distribution - which is quite wrong.
This large $\chi^2$ value arises since in this case
the average value of $\chi^2$ involves the sum of the inverses
of the eigenvalues of $C$. A very small eigenvalue will have the
effect of increasing $\chi^2$ more than that obtained from
the estimated value of 1 for each eigenvalue. Again this effect
is of $O(1/N)$  as for the correlated case above.
An analytic calculation for large $N$ gives
$\chi^2/D=1+(D+1)/N $.  The difference
is that the coefficient now increases   with $D$.
Moreover the expression becomes infinite at $N=D$.

This effect
can be even more pronounced in making a constrained fit
since the distortion of the surface may shift the best fit
value as well as affecting the probability of the fit.
We simulated such a fit by finding the most probable
point such that
$$
X_i = F_i(a) =  a \delta_{i1}
$$
This is a one parameter fit and the true value of $a$ is zero.
For each sample, we find the $a$ value giving the maximum
probability $Q_G$. We can also estimate the error on
this best fit parameter within each sample by looking at
the second derivative of $\chi^2$ with respect to $a$
in the usual way~\cite{T}.
For $N=10$ and $D=5$ for a correlated fit,
we find the error on the parameter $a$ to be
0.056 on average over many samples.  However, the average
over many samples of the best fit value of $a$ itself is $<a^2>=0.20$.
Thus, in this case,  the error ascribed
 to the best fit value from the study of the quadratic sample
surface is too small -
 about one half of that actually present. This is a warning
that the severe distortion of the sample gaussian surface can
upset the naive error analysis.

For the same situation with $N=10$ and $D=5$ but an uncorrelated fit, we
get $<a^2>=0.10$ both from the second derivative
error analysis within each sample and also from the
observed distribution of the best fit over many samples.
Thus a diagonal $\chi^2$ fit gives quite consistent results.
Moreover the true error on $a$ is larger from a correlated fit.
This can be understood since the diagonal $\chi^2$ fit makes a
stronger assumption about the nature of the data set and so
allows a tighter fit.

Furthermore the acceptability of these fits can be
obtained from  the
value of $\chi^2/(D-1)$ itself. We find a  much bigger value for a
correlated fit (2.26) than for a diagonal fit (1.28). At face
value, this would suggest that the correlated fit was unacceptable.
This latter effect is just the same as that described before and
is due to the bias from the sample fluctuations at small $N$.

A standard method of finding the error on a sample value
of  fit parameters $a$ is to repeat the whole fit with many
bootstrap samples derived from the original sample.
This corresponds to using samples of the point
distribution $Q_S(X)$ introduced earlier.
 A small bias
can be introduced by this procedure and this can be corrected
as discussed in Appendix A. With
the correlated $\chi^2$ method, this is inherently
dangerous since the fluctuations in the shape of the gaussian
surface can be exaggerated. For example if $N=10$ and $D=5$,
a bootstrap sample of $N$ may only have contributions from
5 data sets. Thus the smallest eigenvalue will be zero with
drastic consequences. In contrast, the bootstrap method
is a completely satisfactory method for finding the
overall errors with a diagonal $\chi^2$ fit. A jackknife
method can also be used instead of bootstrap but it has
no obvious advantages except less computation.

Thus we learn that a representation of the probability
surface as a general gaussian is rather unstable if
the number of data points is insufficient. In general we
need $N > D^2 $ to have a reasonable description with
no big distortion of the shape in any direction. This
can be understood roughly from the fact that the
$D$-dimensional surface is determined by $D(D+1)/2$ real
numbers. The distortion averaged over all directions
is less than that in the worst direction. Thus the
$\chi^2$ estimates made above need $N > 10 (D+1)$ to
be less that $ 10\%$. To avoid serious fluctuations
in sample estimates, it is prudent to take
$N > \max( D^2, 10(D+1))$.

If one has fewer data points than this, then either one can
estimate the expected probability distributions by simulation as
in the example above, or one can assume a simpler model of the
surface. The simplest model is to treat the data as uncorrelated.
Then one has effectively $D$ one dimensional situations. Put
another way, the minimum width is now to be  taken along the
prescribed $D$ axes not along any arbitrary direction.
For example we use the same distribution introduced
above but now with the different dimensions treated as
uncorrelated. Then with $D=7$,  $N=13$ data points is sufficient
to have the smallest  eigenvalue greater than 0.5 (to be
compared with $N=55$ needed for correlated $\chi^2$).
Thus
the shape is much more stable to fluctuations. Of course this
approach is only acceptable if there are indeed no
 statistically-significant correlations
in the data - as is the case in our example distribution.

 A more realistic approach is to model such correlations
by less parameters than a general quadratic - for instance an
overall common magnitude may represent the main effect.
So putting $y_i=x_i/x_1$ with $y_1=x_1$ may result in
$y$ being less correlated than $x$. Such an approach
 has the side effect that $y_i$ are no longer unbiased
estimators. To proceed, one makes a diagonal $\chi^2$ fit
to the data sample in the $y$ variables. The error in the
fit parameters is taken from a bootstrap analysis in which
the original $x$ data are bootstrapped. Then any bias can be
accommodated as shown in the appendix.

\section{Conclusion}

Don't use correlated $\chi^2$ with $N$ data samples of $D$
data unless $N> \max(D^2,10(D+1))$.  If you have insufficient data
samples but there is statistically-significant evidence for correlation
among the data, then try to model any correlations among the data
with less parameters. Even if some correlation among the data is
suspected, it is reasonable  to use
an uncorrelated $\chi^2$ fit but to estimate the errors
on the parameters by an overall bootstrap of the fit rather than
from the dependence of $\chi^2$ on the parameters. The
only drawback of such a procedure is that it may be
difficult to estimate the goodness of fit since the
correlations among the data may not have been adequately
treated.

\section{Acknowledgements}

I thank my colleagues in the UKQCD Collaboration for
helpful discussions.

\appendix

\section{ Bootstrap biases}

Here we discuss biases that can result from evaluating the
average of some function of a quantity which is an unbiased
estimator. We define an unbiased estimator as one for
which the sample value $\overline x$ averaged over many
samples is the true average $\hat x$.
So $<\overline x> = \hat x$. We have in mind that $x$ will
be the sample average of some unbiased data and its distribution
$Q_S$ can be estimated by bootstrap as described in section 2.

Now if $x$ is unbiased then $f(x)$ will in
general be biased. A simple illustration that explains
the origin of this effect is that if $x$ has a  gaussian
distribution with standard error $\sigma$ then
$ (x \pm \sigma)^2=x^2 + {\sigma}^2 \pm 2 x \sigma$. Hence the
average of $x^2$ is shifted systematically from the (average
of $x$ )$^2$.

If the distribution of $x$ is $Q_S(x)$ with sample average $\overline x$,
 then  the average of $f$ over this distribution is
$$
\overline{f} = \int dx f(x) Q_S(x) \approx f(\overline{x})
+ {1 \over 2} v f''(\overline{x})
$$
where $v$  is the variance of the distribution $Q_S$. This
is a rough estimate of the bias. Now the distribution $Q_S(x)$
is itself only a sample and it has a sample mean $\overline x$
which is distributed (in an unbiased way) around the
true mean $\hat x$ with distribution $Q_T$ with
variance $v_T$. To $O(1/N)$, the true variance $v_T$ is equal to
the sample variance $v$.   The average over
many samples gives
$$
{<f(\overline x)>} = \int dx f( \overline x) Q_T (\overline x -\hat x)
 \approx f(\hat x)+ {1 \over 2} v f''(\hat x)
$$

Thus there is a further bias from the distribution of the sample  mean.
The key is that this bias is the same (to order $1/N$)  as that above:
namely the difference between the sample values of $\overline f$ and
$f(\overline x)$. This suggests the unbiased  estimator for $f(\hat x)$:
 $$
 f_U =  2 f(\overline x) - \overline f
 $$
 Now this expression
can be evaluated in a straightforward way  by bootstrap simulation. The
average of $f(\overline x)$ over the bootstrap   samples gives
 $\overline f$, while $f(\overline x)$ is obtained
by evaluating $f$ with the sample average  $\overline x$. Note that since
$x$ is an unbiased estimator, the average of $\overline x$ over
the bootstrap samples should be the same as the sample average in the
limit of many bootstrap samples.

This technique is also valid for the parameters $a$ of a fit to the data.
The combination $2a-<a>_B$ will be unbiased where  the angle
brackets here imply an average of the $a$ values obtained
from fitting  many bootstrap samples of the original  data set.

The unbiased estimator can be thought of as the naive value
$f(\overline x)$ with a systematic correction
$( f(\overline x)-\overline f)$. This correction will itself have
some error when evaluated from one data sample.
An estimate of the error to be ascribed to this unbiased estimator
can be obtained by a further bootstrap. One creates many first
level bootstrap samples, and for each one calculates $f_U$ by a
further nested bootstrap. Then the variance of the values
of $f_U$ obtained is available. This is computationally
demanding but straightforward in practice.
This is analogous
to the nested jackknife method proposed to deal with this same
problem of bias~\cite{B}.

\newpage

\begin{table}
\begin{center}
\caption{Correlated $\chi^2/D$ for $N$ data   with dimension $D$
(averages from 10000 samples)}
\vskip 5mm
\label{tabletwo}

\begin{tabular}{lllllll} \hline
  $N$  & $D=1$ & $D=3$ & $D=5$ & $D=7$ & $D=10$ & $D=15$ \\ \hline
  $10 $ &  $1.29$ & $1.83$ &  $3.03$ & $9.16$ & $ \infty$ & $\infty$ \\
  $20 $ &  $1.12$ & $1.25$ &  $1.44$ & $1.74$ &  $2.38$ & $6.45$ \\
  $30 $ &  $1.07$ & $1.15$ &  $1.27$ & $1.38$ &  $1.59$ & $2.22$ \\
  $40 $ &  $1.05$ & $1.13$ &  $1.17$ & $1.25$ &  $1.39$ & $1.70$ \\
  $50 $ &  $1.04$ & $1.08$ &  $1.13$ & $1.21$ &  $1.30$ & $1.49$ \\
 $100 $ &  $1.02$ & $1.03$ &  $1.06$ & $1.09$ &  $1.12$ & $1.19$ \\
      \hline
	       \end{tabular}
	      \end{center}
	       \end{table}

\begin{table}
\begin{center}
\caption{Data points $N$ needed for smallest eigenvalue
 0.5 with dimension $D$}
\vskip 5mm
\label{tableone}

\begin{tabular}{ll} \hline
 $D$ & $N$     \\ \hline
 $3$ & $16 $  \\
 $5$ & $35 $  \\
 $7$ & $55 $  \\
 $10$ & $86 $  \\
 $15$ & $140 $  \\
 $20$ & $196 $  \\        \hline
	       \end{tabular}
	      \end{center}
	       \end{table}

\newpage

\begin{figure}[h]
\centerline{\psfig{figure=fluct.ps,height=6in}}
\caption{ The largest and smallest eigenvalues of $C$ for
$N$ data samples in $D$ dimensions}
\end{figure}


\begin{thebibliography} {99}

\bibitem{T} D. Toussaint, {\it From Actions to Answers},
World Scientific 1990, (eds. T. DeGrand and D. Toussaint) p. 121.

\bibitem{P} Private communication: UKQCD and many other lattice groups.

\bibitem{S} D. Seibert, CERN preprint TH.6892/93.

\bibitem{B} B. A. Berg, FSU preprint FSU-SCRI-90-100.

\end{thebibliography}
\end{document}